\documentclass[epj,fleqn,referee]{svjour}

\usepackage{enumerate,amsmath,bbm,graphicx}

\newcommand {\ie}{{i.e., \/}}
\newcommand {\eg}{{e.g., \/}}
\newcommand {\etal}{{et al. \/}}

\newcommand{\mat}[1]{\vec{#1}}

\def\A{{\cal A}}
\def\un{\mathbbm{1}}

\begin{document}



\title{Mixing Bandt--Pompe  and Lempel--Ziv  approaches: another way  to analyze
  the complexity of continuous-state sequences}

\titlerunning{Mixing  Bandt--Pompe  \& Lempel--Ziv  approaches:  another way  to
  analyze the complexity\ldots}


\author{S. Zozor\inst{1,2}  \and D. Mateos\inst{3} \and  P. W. Lamberti\inst{3}}
\institute{  Laboratoire  Grenoblois  d'Image,  Parole,  Signal  et  Automatique
  (GIPSA-Lab), CNRS, et Universit\'e de Grenoble, 961 rue de la Houille Blanche,
  38402           Saint            Martin           d'H\`eres,           France,
  \email{steeve.zozor@gipsa-lab.grenoble-inp.fr} \and
  Instituto  de F\'{\i}sica  de La  Plata (IFLP),  CONICET, and  Departamento de
  F\'{\i}sica, Facultad  de Ciencias Exactas, Universidad Nacional  de La Plata,
  C.C.~67, 1900 La Plata, Argentina, \email{szozor@fisica.unlp.edu.ar} \and
  Facultad  de  Matem\'atica,  Astronom\'ia  y F\'isica  (FaMAF),  CONICET,  and
  Universidad   Nacional   de  C\'ordoba,   Avenidad   Medina  Allende,   Ciudad
  Universitaria,           CP:X5000HUA,           C\'ordoba,           Argentina
  \email{mateos@famaf.unc.edu.ar, lamberti@famaf.unc.edu.ar}}


\abstract{ In this paper, we propose to mix the approach underlying Bandt--Pompe
  permutation  entropy  with Lempel--Ziv  complexity,  to  design  what we  call
  Lempel--Ziv permutation  complexity. The principle consists of  two steps: (i)
  transformation of a continuous-state series that is intrinsically multivariate
  or arises  from embedding  into a sequence  of permutation vectors,  where the
  components  are the positions  of the  components of  the initial  vector when
  re-arranged;  (ii) performing the  Lempel--Ziv complexity  for this  series of
  `symbols', as  part of a discrete  finite-size alphabet. On the  one hand, the
  permutation entropy of Bandt--Pompe aims at the study of the entropy of such a
  sequence; \ie  the entropy of patterns  in a sequence (\eg  local increases or
  decreases).  On the other hand, the Lempel--Ziv complexity of a discrete-state
  sequence aims  at the study of  the temporal organization of  the symbols (\ie
  the  rate  of  compressibility   of  the  sequence).   Thus,  the  Lempel--Ziv
  permutation complexity aims  to take advantage of both  of these methods.  The
  potential from  such a combined approach  -- of a permutation  procedure and a
  complexity analysis -- is evaluated through the illustration of some simulated
  data and some real data.  In  both cases, we compare the individual approaches
  and the combined approach.\vspace{-5mm}
  \vspace{-5mm}
}

\maketitle


\section{Introduction}
\label{introduction:sec} 

Many real signals result from very complex dynamics and/or from coupled dynamics
of many dimensional systems.  Various examples  can be found in biology, such as
the reaction--diffusion process in  cardiac electrical propagation that provides
electrocardiograms, and  the collective actions  of genes for the  production of
proteins  in  specific  quantities~\cite{ZozBla03,Kau93,BotSmi95}.  In  finance,
there is the  example of the variation  in the price of an  asset, which results
from  the collective actions  of the  buyers and  sellers~\cite{FerFra03}, while
statistical physics  and social  sciences also have  huge numbers  of situations
where `complexity'  emerges~\cite{Art94}.  One of the challenges  is to describe
these  complex  signals  in a  simple  way,  to  allow meaningful  and  relevant
information to be extracted~\cite{Raj00,RajMih03,CysBet00,QuiArn00,ZhaZhu99}.

The complex origin of such signals  has led researchers to analyze these signals
through tools that come either from the `probability world', or conversely, from
`nonlinear dynamics'.  The purpose is  to characterize the degree of information
or the complexity of the signals under analysis as well as possible.
The first approach is statistical, and the  goal is to measure the spread of the
distribution underlying  the data, or to  detect any changes  in the statistics.
The    common   tools    that   are    used   here    come    from   information
theory~\cite{CysBet00,QuiArn00,DarWue00,EbeMol00,TorGam00},  or  are correlation
measures~\cite{BotSmi95}, or come from spectral analysis~\cite{EvrMoo01}.
The second  approach is  devoted to signals  that are produced  by deterministic
(generally nonlinear) mechanisms, even if the sequence under analysis can appear
to be somewhat  `random'.  The tools generally used for  the description of such
complex  signals come  often  from  the chaos  world,  like Lyapunov  exponents,
fractal dimensions,  and others~\cite{Raj00}, or from the  concept of complexity
in       the       sense        of       Kolmogorov       (\eg       Lempel--Ziv
complexity)~\cite{RajMih03,ZhaZhu99,Rad02,SzcAmi03,ZozRav05}.

The measures  from information theory  are very powerful,  in a sense  that they
allow  the quantification  of a  degree of  uncertainty (the  rate) of  a random
sequence, or  of a  sequence considered as  randomly generated.   However, tools
such as entropies  can have some drawbacks when used in  practice.  One of these
occurs when dealing with continuous-state data.  In this case, the estimation of
a   differential   entropy    from   the   data   is   not    always   an   easy
task~\cite{BeiDud97,LeoPro08,AmbZoz08}.  Some  nonparametric estimators make use
of  nearest  neighbors, or  of  graph  lengths,  although their  properties  are
difficult   to   study~\cite{BeiDud97,LeoPro08,AmbZoz08,SchGra96,Sch04}.    More
simple estimators are based on `plug-in' approaches~\cite{BeiDud97}; namely, the
density is estimated  using a Parzen-Rosenblatt approach~\cite{Ros56,Par62}, and
the estimation is plugged into  the mathematical expression of the entropy.  The
most simple  density estimator is based  on a histogram, which  is equivalent to
quantization  of   the  data.   The  estimation  performance   depends  on  this
quantization (\eg  number of  thresholds, quantization intervals).   To overcome
this  potential  difficulty, Bandt  and  Pompe  proposed  (i) to  construct  the
multivariate trajectories from the scalar  series, \ie an embedding; and (ii) to
work  with the  so-called vectors  of  permutation, \ie  for each  point of  the
trajectory,  its components  are  sorted, and  each  component of  the point  is
replaced by  its position  (rank) in the  rearranged components~\cite{BanPom02}.
Bandt  and  Pompe  proposed  then  to  estimate  the  discrete  entropy  of  the
permutation  vector  sequence,  which  led  to the  so-called  {\em  permutation
  entropy},     and    later     on,    to     some    variations     of    this
measure~\cite{KelSin05,BiaQin12,FadChe13}.  However, when dealing with sequences
generated  by  a  deterministic   process,  such  statistical  measures  can  be
inappropriate because they measure an ensemble, or average, behavior.

Conversely, for  deterministic sequences  generated by dynamical  systems, there
are  a huge  number  of analysis  tools,  like Lyapunov  exponents, and  fractal
dimensions,  among  others~\cite{GraPro83:10,LasMac94,Rob11}.   In general,  the
quantities under  study are relatively  difficult to evaluate, and  they require
long times of computation.  As an  example, there can be the need to reconstruct
a phase-space  trajectory using several  estimations to determine  the embedding
dimension and  the optimal delay, and then,  in a second step,  to estimate some
quantities  from  the  reconstructed  trajectory,  such as  the  whole  Lyapunov
spectrum,    or    just    some    exponents    (\eg    positive,    max),    or
dimensions~\cite{Tak81,Rob11}.   Moreover, these  tools  are generally  designed
specifically  for  the study  of  chaotic series.   A  more  natural concept  of
`uncertainty'  of  a  time series,  whether  chaotic  or  not,  is that  of  its
complexity in the sense of Kolmogorov.  Roughly speaking, this measures the {\em
  minimal size  of a  binary program  that can generate  the sequence}  (\ie the
algorithmic  complexity)~\cite{CovTho06,GacTro01}.  Among  these,  there is  the
Lempel--Ziv  complexity,   which  is  based  on   simple  recursive  copy--paste
operations,  as  will  be  seen later~\cite{LemZiv76,ZivLem77}.   This  kind  of
measure     naturally     finds      applications     in     the     compression
domain~\cite{CovTho06,ZivLem77,Wel84},   and  it   is  also   used   for  signal
analysis~\cite{ZhaZhu99,TorGam00,Rad02,SzcAmi03}.  A strength of this complexity
is that as it deals with  a random discrete-state and ergodic sequence, and when
it  is  correctly   normalized,  it  converges  to  the   entropy  rate  of  the
sequence~\cite{LemZiv76,CovTho06}.   In  a  sense,  the  Lempel--Ziv  complexity
contains the concept of complexity  both in the deterministic sense (Kolmogorov)
and in  the statistical sense  (Shannon).  This property  led to the use  of the
Lempel--Ziv complexity for entropy estimation purposes~\cite{Han89,SchGra96}.  A
possible  drawback of  the  Lempel--Ziv complexity  is  that it  is defined  for
sequences that take  their values on a discrete (finite  sized) alphabet.  If it
can find natural  applications that deal with discrete-state  sequences, such as
DNA  sequences or  sequences generated  by logical  circuits,  while `real-life'
signals are generally continuous states\footnote{When performing the acquisition
  of  a  signal in  a  computer,  for example,  the  (discrete  time) series  is
  intrinsically  a discrete-state  series due  to  the finite  precision of  the
  computer. However, this precision is generally high, so that the series can be
  assumed  to be  a continuous-state  series.   In particular,  in general,  the
  number of  possible states  is much higher  that the  number of samples  to be
  analyzed.}.    Thus,   to   use   the  Lempel--Ziv   complexity   for   signal
characterization purposes, there  is first the need to  quantize the data, which
introduces some parameters into the  tuning.  These parameters can influence the
behavior  of  the complexity  of  the  quantized signal,  as  can  be seen,  \eg
in~\cite{KasSch87},  where  for  a  logistic  map,  some  bifurcations  are  not
(completely) captured by the Lempel--Ziv complexity.

As can  be imagined,  there are many  ways to  overcome the drawbacks  of purely
statistical methods or purely deterministic approaches.  Here, we concentrate on
the Lempel--Ziv complexity, using first the idea that underlies the Bandt--Pompe
entropy, to `quantize' a sequence to analyze.

This report is organized as follows. 
In  section~\ref{Recalls:sec},  we first  define  the  notation  we use  in  the
following  sections. Then  we provide  some basics  on Bandt--Pompe  entropy (or
permutation  entropy).  In  the same  section, we  also provide  some  basics on
Lempel--Ziv  complexity, proposing  then to  `mix' both  of these  approaches in
section~\ref{BPLZ:sec}, to  give what we  call the {\em  Lempel--Ziv permutation
  complexity}.   In  this  same  section,  we provide  some  properties  of  the
Lempel--Ziv  permutation  complexity, including  in  an  Appendix the  technical
details and the description of a practical way to calculate this complexity when
dealing     with     scalar     sequences.      We    then     illustrate     in
section~\ref{Illustrations:sec} how  the Lempel--Ziv permutation  complexity can
be used  for data analysis of  both simulated sequences  and biological signals,
and we finish the paper by drawing up our concluding remarks.


\section{Notation and recall}
\label{Recalls:sec}


\subsection{Bandt--Pompe permutation entropy}

The starting point of  the Bandt--Pompe approach~\cite{BanPom02} appears to take
its origin  from a  study of  chaos, and more  specifically, through  the famous
Takens'  delay  embedding  theorem~\cite{Tak81,Rob11}.   The principle  of  this
theorem is the reconstruction of the state trajectory of a dynamical system from
the observation of one of its  states.  To fix the ideas, consider a real-valued
discrete-time series $\{  X_t \}_{t \ge 0}$ that  is assumed to be a  state of a
multidimensional trajectory.  Consider two integers  $d \ge 2$ and $\tau \ge 1$,
and from  the series,  let us  then define a  trajectory in  the $d-$dimensional
space as:
\begin{equation}
\vec{Y}^{(d,\tau)}_t = \left[ X_{t-(d-1)\tau} \quad \cdots \quad X_{t-\tau} \: X_t
\right]^t, \quad t \ge (d-1) \tau
\end{equation} 
where the  dimension $d$ is  known as the  {\em embedding dimension},  and where
$\tau$ is called  the {\em delay}.  Takens' theorem gives  conditions on $d$ and
$\tau$ such  that $\vec{Y}^{(d,\tau)}_t$  preserves the dynamical  properties of
the     full     dynamic    system     (\eg     reconstruction    of     strange
attractors)~\cite{Tak81,Rob11}.    Many  studies   have  dealt   with  `optimal'
reconstruction  of this phase  space; \ie  the choice  of the  correct embedding
dimension, and more particularly, the `optimal' delay.

In~\cite{BanPom02}, Bandt and Pompe did not focus especially on chaotic signals,
even if these  signals serve as illustrations.  Thus, they did  not focus on the
phase-space  reconstruction  problem.   More  precisely, they  did  not  provide
discussion on the parameters $d$ and $\tau$.  The only ingredient they wished to
conserve  was  the idea  of  taking  into account  the  dynamics  of the  system
underlying an  observed signal.  These questions of  optimal reconstruction also
go  beyond the  scope of  our paper,  so we  do not  discuss the  choice  of the
embedding dimension and of the delay in the sequel anymore.

Starting  with  the phase-space  trajectory  $\vec{Y}^{(d,\tau)}_t$, instead  of
focusing  on the real-valued  vectors, Bandt  and Pompe  were interested  in the
order of  the components of  the vectors.  The  principle consists first  of the
sorting (in  ascending order) of  the components of  $\vec{Y}^{(d,\tau)}_t$, and
then the replacement  of each component $X_{t-k \tau}$ by  its rank/ position in
the  sorted vector.   This  so-called  {\em permutation  vector}  is denoted  as
$\vec{\Pi}_{\vec{Y}_t^{(d,\tau)}}$  in  the following.   As  an  example, for  a
vector $\vec{Y} = [Y_0 \quad Y_1 \quad  Y_2]^t$ such that $Y_2 \le Y_0 \le Y_1$,
the permutation vector is $\vec{\Pi}_{\vec{Y}} = [1 \quad 2 \quad 0]^t$. Dealing
with  random processes,  it  is then  possible  to define  the {\em  permutation
  entropy} as the Shannon entropy $H$ of the (random) permutation vector
\begin{equation}
H^{\mathrm{\pi}}_{d,\tau}(X_t) \equiv H\left( \Pi_{\vec{Y}_t^{(d,\tau)}} \right)
\label{PermutationEntropy:eq}
\end{equation}
For a stationary  process, provided the size of the sequence  is large enough in
terms of $d!$, the entropy can be estimated via the frequencies of occurrence of
any   of    the   $d!$   possible   permutation   vectors    in   the   sequence
$\vec{Y}^{d,\tau}_t$.  In  their paper, Bandt and Pompe  defined the permutation
entropy  as  the   Shannon  entropy  of  the  frequencies   of  the  permutation
vectors\footnote{More  precisely,  in their  paper,  the  permutation vector  is
  defined as the time position of the component in the sorted vector, instead of
  the vector  of the rank  of the vector  components.  As there is  a one-to-one
  mapping between the two ways of making,  the entropy of the two vectors is the
  same.},  which  gives asymptotically  the  entropy $H^{\pi}_{d,\tau}(X_t)$  of
Equation~\eqref{PermutationEntropy:eq} when dealing  with a long-time (infinite)
stationary and ergodic process,  as indicated in~\cite{BanPom02}.  Starting from
a  sequence  of  length  T,  $X_0  \ldots  X_{T-1}$,  in  the  sequel  we  write
$\widehat{H}^{\pi}_{d,\tau}(X_{0:T-1})$ for  the entropy of  the frequencies, to
distinguish this from the entropy of the random process.  Several quantifiers of
information  based   on  $\widehat{H}^{\pi}_{d,\tau}(X_{0:T-1})$  were  proposed
in~\cite{BanPom02},  although  such extensions  go  beyond  the  purpose of  the
present paper. Thus, we do not present these here.

The  idea behind  permutation  entropy  is that  the  $d!$ possible  permutation
vectors,  also called {\em  patterns}, might  not have  the same  probability of
occurrence,  and  thus,  this  probability  might  unveil  knowledge  about  the
underlying system.   For a sequence  of independent and  identically distributed
(iid) variables,  whatever the distribution of  the random variable,  all of the
patterns  have the  same probability  $\frac{1}{d!}$ of  occuring  (whatever the
delay  $\tau$),  so  that  the  permutation  entropy is  maximum  and  equal  to
$\log(d!)$~\cite{BanPom02}. Conversely, an important situation is represented by
the so-called {\em forbidden patterns}, which are patterns that do not appear at
all in the analyzed  time series~\cite{AmiKoc06,AmiZam07,Ami10}.  As an example,
it was  shown in the logistic  map $X_{t+1} =  4 X_t (1-X_t)$ that  whatever the
initialization $X_0$,  for $d =  3$ and $\tau  = 1$, the permutation  vector $[2
\quad  1  \quad  0]^t$ never  appears.   Such  behavior  shows  how the  use  of
permutation vectors  allows the  distinguishing between purely  random sequences
and deterministic sequences (\eg when the  last one is chaotic, and thus appears
random): some  authors have said that  the presence of forbidden  patterns is an
indicator   of   deterministic  dynamics~\cite{AmiKoc06,AmiZam07,Ami10}.    This
question remains, however, controversial, as  it is possible to construct random
series with forbidden patterns~\cite{RosCar12}, and conversely, a chaotic series
does not always show forbidden patterns~\cite{RosOli13}.

Note that if we work on a multidimensional sequence $\{ \vec{X}_t \}_{t \ge 0}$,
the permutation procedure  can be performed on each  vector $\vec{X}_t$, so that
there are no  embedding procedures.  To distinguish this  situation from that of
Bandt and Pompe,  we denote the permutation entropy and  its estimate as $H^\pi$
and $\widehat{H}^\pi$, respectively, without  mention of any delay and embedding
dimension.


\subsection{Lempel--Ziv complexity}

Consider a finite-size sequence $S_{0:T-1} = S_0 \ldots S_{T-1}$ of symbols that
take  their values  in  an alphabet  $\A$ of  finite  size $\alpha  = \left|  \A
\right|$.  In 1965, Kolmogorov introduced  the concept of the complexity of such
a  sequence as the  size of  the smallest  binary program  that can  produce the
sequence~\cite{CovTho06}.   In an algorithmic  sense, the  Kolmogorov complexity
measures the  minimal `information'  contained in the  sequence, or  the minimal
information needed to  generate the sequence.  Several years  later, the seminal
work of Lempel and Ziv appeared~\cite{LemZiv76}, which dealt with the complexity
of the Kolmogorov type of a sequence, restricting this concept to the `programs'
based  only on  two  operations:  recursive copy  and  paste operations.   Their
definition lies in the two fundamental concepts of reproduction and production:
\begin{itemize}
\item {\bf Reproduction:}  this consists of extending a  sequence $S_{0:T-1}$ of
  length  $T$,   adding  a   sequence  $Q_{0:N-1}$  via   recursive  copy--paste
  operations, which  leads to  $S_{0:T+N-1}$, \ie the  first letter $Q_0$  is in
  $S_{0:T-1}$, let us  say $Q_0 = S_i$,  the second one is the  following one in
  the extended sequence of size $T+1$, \ie $Q_1 = S_{i+1}$, etc~: $Q_{0:N-1}$ is
  a subsequence of  $S_{0:T+N-2}$.  In a sense, all of  the `information' of the
  extended sequence $S_{0:T+N-1}$ is in $S_{0:T-1}$.
\item {\bf  Production:} the  extended sequence $S_{0:T+N-1}$  is now  such that
  $S_{0:T+N-2}$  can be  reproduced  by  $S_{0:T-1}$.  The  last  symbol of  the
  extension can  also follow  the recursive copy--paste  operation, so  that the
  production is a reproduction, but can be `new'.  Note thus that a reproduction
  is a production, but the converse is false.
\end{itemize}
Any sequence can  be viewed as constructed through  a succession of productions,
called a history.  As an example, a sequence can be `produced' symbol by symbol.
However, a given  sequence does not have a unique  history; several processes of
productions can  lead to  the same  sequence.  In the  spirit of  the Kolmogorov
complexity,  Lempel and  Ziv were  interested in  the optimal  history;  \ie the
minimal productions  needed to generate  a sequence~: the  so-called Lempel--Ziv
complexity, denoted as  $C(S_{0:T-1})$ in the following, is  this minimal number
of production  steps needed for the  generation of $S_{0:T-1}$. In  a sense, $C$
describes the `minimal' information needed to generate the sequence by recursive
copy--paste operations.   Thus, the approach of  Lempel and Ziv,  and of several
variations~\cite{ZivLem77,Wel84}, naturally  gave rise to  various algorithms of
compression  (including the famous  `gzip').  It  can intuitively  be understood
that  in a  minimal  sequence of  production,  all of  the  productions are  not
reproductions,  otherwise  it  would  be   possible  to  reduce  the  number  of
steps~\cite{LemZiv76}.  This  allowed the  development of simple  algorithms for
the evaluation of the Lempel--Ziv complexity of a sequence~\cite{KasSch87}.

Surprisingly,  although analyzing  a  sequence from  a completely  deterministic
point  of view,  it  appears  that $C(S_{0:T-1})$  sometimes  also contains  the
concept  of  information   in  a  statistical  sense.   Indeed,   it  was  shown
in~\cite{CovTho06,LemZiv76} that  for a  random stationary and  ergodic process,
when correctly normalized,  the Lempel--Ziv complexity of the  sequence tends to
the entropy rate of the process; \ie
\begin{equation}
\lim_{T \to + \infty} C(S_{0:T-1}) \frac{\log(T)}{T} \: = \: \lim_{T \to +
\infty} \frac{H(S_{0:T-1})}{T}
\label{CLZ_ER:eq}
\end{equation}
where $H(S_{0:T-1})$ is the joint entropy  of the $T$ symbols, and the righthand
side is the  entropy rate (entropy per symbol) of the  process.  Such a property
gave  rise to  the  use of  the  Lempel--Ziv complexity  for entropy  estimation
purposes~\cite{Han89,SchGra96}.

Note that  using the Lempel--Ziv complexity  for analysis purposes  might not be
envisaged if the size  of the sequence is not large enough  in terms of the size
of  the alphabet.   Indeed, for  small  sequences compared  to the  size of  the
alphabet, except for very  elementary situations (\eg constant signals, periodic
signals), the complexity of the sequence  has a great probability of being close
to the size of the sequence.


\section{The Lempel--Ziv permutation complexity}
\label{BPLZ:sec}

As we have just  seen, in a sense, the Lempel--Ziv complexity  aims to capture a
level  of redundancy,  or of  regularity,  in a  sequence.  Thus,  this tool  is
interesting for the analysis of signals  that appear to be random, but that hide
some  regularities, such as  in chaotic  sequences~\cite{KasSch87}.  Conversely,
viewing this complexity as an estimator of the Shannon entropy when dealing with
random sequences, its use is also relevant  in such a context. In some sense, it
provides  a  bridge  between  the  two  above-mentioned  contexts.   However,  a
disadvantage  of the  Lempel--Ziv  complexity is  that  it is  defined only  for
sequences of symbols  taken in a discrete (finite  size) alphabet.  Dealing with
`real-life' sequences, a quantization has to be performed before its use, as has
been  done  in  many  of  the  studies  dealing  with  data  analysis  via  this
complexity~\cite{ZhaZhu99,Rad02,SzcAmi03}.  Quantizing a  signal might have some
consequences  in  the evaluation  of  the complexity,  and  the  effects of  the
parameters of the quantizers appear difficult to evaluate.

Conversely,  the  permutation  entropy  also  has  some  drawbacks  due  to  its
statistical aspects.  To illustrate why sometimes it cannot capture the dynamics
of a sequence,  consider the example of an iid scalar  noise, versus a periodic
scalar sequence  of period $T  = 2$. For  an embedding dimension  $d = 2$  and a
delay $\tau = 1$, in both cases  the permutation vectors $[0 \quad 1]^t$ and $[1
\quad 0]^t$ appear with the same frequency $\frac12$ (assuming the length of the
sequence  is large  enough).  Thus,  the permutation  entropy is  equal  in both
cases,  and in  this example  it is  thus not  sensitive enough  to discriminate
between  the  random  iid   sequence  and  the  periodic  sequence\footnote{More
  rigorously, it is known that  using the permutation entropy for data analysis,
  several embedding  dimensions have to be  tested.  For $d=3$  in this example,
  the permutation  entropy makes the distinction  between the iid  noise and the
  periodic  sequence.}.   Several variants  to  avoid  such  a drawback  can  be
imagined;  \eg  taking  into  account  the  amplitudes  when  constructing  the
permutation     vectors.     The    weighted-permutation     entropy    proposed
in~\cite{FadChe13} shows its efficiency for the detection of abrupt changes in a
sequence, but  in the example given above,  it will not be  able to discriminate
between  the   two  situations.   Moreover,  when  dealing   with  an  intrinsic
multidimensional sequence,  the permutation vectors  do not clearly  reflect any
dynamics.

\

To avoid the possible disadvantages of  both methods, we propose here to mix the
Bandt--Pompe  and  Lempel--Ziv  approaches;  \ie  to  analyze  the  sequence  of
permutation vectors via the Lempel--Ziv complexity.  In this way, it is expected
that we  can take advantage  of both methods,  and thus reduce  their respective
drawbacks.   In  the  following,  the  so called  {\em  Lempel--Ziv  permutation
  complexity} of a finite length  scalar sequence $X_{0:T-1}$ or a finite length
multivariate sequence $\vec{X}_{0:T-1}$ are respectively denoted as:
\begin{equation}
C_{d,\tau}^\pi(X_{0:T-1}) \equiv C\left( \vec{\Pi}_{\vec{Y}_{(d-1)
\tau}^{(d,\tau)}}\ldots \vec{\Pi}_{\vec{Y}_{T-1}^{(d,\tau)}}\right)
\end{equation}
where $\vec{Y}_t^{(d,\tau)}  = [X_{t-(d-1)  \tau} \quad \ldots  \quad X_{t-\tau}
\quad  X_t]^t$  and  $\vec{\Pi}_{\vec{Y}_{T-1}^{(d,\tau)}}$ is  its  permutation
vector, and
\begin{equation}
C^\pi(\vec{X}_{0:T-1}) \equiv C\left( \vec{\Pi}_{\vec{X}_0} \ldots
\vec{\Pi}_{\vec{X}_{T-1}} \right)
\end{equation}

This way  provides an answer  to the necessity  of working with data  taking the
values  on a  finite size  alphabet (here,  the alphabet  is $\A  \equiv \left\{
  \left[ \pi(0) \quad \ldots \quad  \pi(d-1) \right]^t: \pi \in \Pi^{(d)} \right
\}$  of size  $\alpha  = d!$,  where $\Pi^{(d)}$  is  the ensemble  of the  $d!$
possible  permutations on  $\{  0 ,  \ldots  , d-1  \}$).   Moreover, viewing  a
permutation vector  as quantization  of the  data, it is  interesting to  draw a
parallel   with   dynamical    quantization;   namely,   of   the   sigma--delta
type~\cite{GerGra92}.  Indeed, dealing with  scalar real-state sequences, in the
case where $\tau =  1$ and $d = 2$, for instance,  the permutation vector is $[0
\quad 1]^t$ if  the signal increases locally, and is  $[1 \quad 0]^t$ otherwise.
In other words, the two  possible permutation vectors quantize the variations of
the signal  in one bit.   Roughly speaking, a  sigma--delta quantizer acts  in a
similar  way\footnote{More rigorously,  it  quantizes the  difference between  a
  sample and  a prediction of  this sample (the  `delta' part) in one  bit.  The
  prediction is  made from  all of  the past samples,  in general  performing an
  integration  or a  summation (the  `sigma'  part).}.  For  $d >  2$, the  same
parallel  should  be   made  in  some  sense  with   the  so-called  multi-stage
sigma--delta quantizers~\cite{ChoWon89}.  This parallel is another motivation to
use permutation vectors  as a way to quantize a  signal.  Moreover, dealing with
intrinsically multivariate  sequences, the permutation vectors can  be viewed as
(vector)  quantization of  the real-valued  vectors; this  scheme does  not need
tuning     parameters,    contrary     to    standard     vector    quantization
schemes~\cite{GerGra92}.

Working  on  the  permutation  vectors   maintains  the  idea  of  studying  the
occurrences of patterns in a sequence.  By analyzing the permutation vectors via
the  Lempel--Ziv  complexity, a  step  is added  because  how  the patterns  are
temporarily organized is analyzed, rather than the frequency of occurrences.  To
stress this, let us come back to the example of the permutation vector sequences
of an  iid noise versus a  periodic sequence of  period $T = 2$.   As previously
explained, the patterns $[0 \quad 1]^t$ and $[1 \quad 0]^t$ appear with the same
frequency in both cases. However,  the difference between the permutation vector
sequences in the two cases is that in the first case, the two patterns appear in
a random  sequence, while in the  second case, they appear  periodically: in the
first case, the Lempel--Ziv complexity is  then high, while it is low (and equal
to 3) in the second case.  With this very elementary example, it can be seen why
the  Lempel--Ziv   permutation  complexity  of  a  sequence   can  provide  more
information on  the dynamics;  \ie by analyzing  how the patterns  are organized
temporarily, not only in terms of the frequency of occurrence.

Moreover,  dealing with  intrinsically multivariate  sequences, the  argument of
capturing  the  dynamics  that underlie  the  sequence  fails,  as there  is  no
embedding prior  to the quantization that  is made with the  construction of the
permutation  vector. At  least  this question  is  not clear.   In essence,  the
Lempel--Ziv complexity will in a way capture the dynamics of such a multivariate
sequence, which  strengthens the interest  for mixing both the  Bandt--Pompe and
Lempel--Ziv approaches in this context.

The  Lempel--Ziv  permutation complexity  has  some  properties  that have  been
inherited from the  standard Lempel--Ziv complexity. The first  is the link with
the permutation entropy. Indeed, for a stationary ergodic process that is scalar
or multivariate, the sequence of permutations remains stationary and ergodic, so
that Equation  \eqref{CLZ_ER:eq} applies to  the Lempel--Ziv complexity  and the
entropy rate of this sequence, which can be written as:
\begin{equation}
\lim_{T \to \infty} C^\pi_{d,\tau}(X_{0:T-1}) \frac{\log T}{T} \: = \: \lim_{T
\to \infty} \frac{H^\pi_{d,\tau}(X_{0:T-1})}{T}
\end{equation}
and
\begin{equation}
\lim_{T \to \infty} C^\pi(\vec{X}_{0:T-1}) \frac{\log T}{T} \: = \: \lim_{T \to
\infty} \frac{H^\pi(\vec{X}_{0:T-1})}{T}
\end{equation}
The second property is the  invariance of the Lempel--Ziv permutation complexity
to a  given permutation applied to the  components of the vector  of the initial
series; \ie for any permutation matrix $\mat{P}$,
\begin{equation}
C^\pi(\mat{P} \vec{X}_0 \ldots \mat{P} \vec{X}_{T-1}) = C^\pi(\vec{X}_0 \ldots
\vec{X}_{T-1})
\end{equation}
In other  words, if a  sequence of vectors  $\vec{X}_t$ is constructed  from $d$
scalar sequences, the choice of the  order of the components does not modify the
value of  the complexity of the  `joint' sequence. This  property arises because
$\vec{\Pi}_{\mat{P} \vec{X}_t}  = \mat{P} \vec{\Pi}_{\vec{X}_t}$  (permuting the
components of  a vector results in  permuting the components  of its permutation
vector),  together  with the  invariance  of  the  Lempel--Ziv complexity  to  a
one-to-one transformation \cite{ZozRav05}.

As shown  by \cite{ZozRav05} for the  Lempel--Ziv complexity, it  is possible to
build measures associated with  the Lempel--Ziv permutation complexity, although
such possible extensions go beyond the scope of the present paper.

Before moving on to put  the Lempel--Ziv permutation complexity into action, let
us just note the following additional choices:
\begin{itemize}
\item  To take  into  account a  finite  resolution in  data  acquisition or  to
  counteract  possible low  noise in  the  data, we  can introduce  a radius  of
  confidence  $\delta$; \ie  if  the absolute  value  of the  difference of  two
  components is  strictly lower  than $\delta$, then  they are considered  to be
  equal.
\item Performing the permutation procedure,  when two components of a vector are
  equal, we chose  the `smallest' one as that with the  lowest index (the oldest
  one in the case of embedding).
\end{itemize}
(see Appendix for more details and further justification).

Once again, note that using  the Lempel--Ziv permutation complexity for analysis
purposes might not be  feasible if the size of the sequence  is not large enough
in terms of the size of the alphabet $d!$.


\section{Illustrations based on synthetic and real data}
\label{Illustrations:sec}


\subsection{Characterizing the logistic map}

To   illustrate  how   the  Lempel--Ziv   permutation  complexity   can  capture
regularities in  a signal, we consider  here the example of  the famous logistic
map
\begin{equation}
X_{t+1} = k \, X_t \, (1-X_t), \qquad t \ge 0, \qquad k \in (0 \: ; \: 4]
\end{equation}
We initialize $X_0$ randomly in $[0 \: ;  \: 1]$ so that the sequence has a real
value in  the interval $[0 \:  ; \: 1]$. This  map has already been  taken as an
illustration by both  Bandt \& Pompe in~\cite{BanPom02}, and  Kaspar \& Schuster
in~\cite{KasSch87}.

\begin{figure}
\centerline{\includegraphics[width=.485\textwidth]{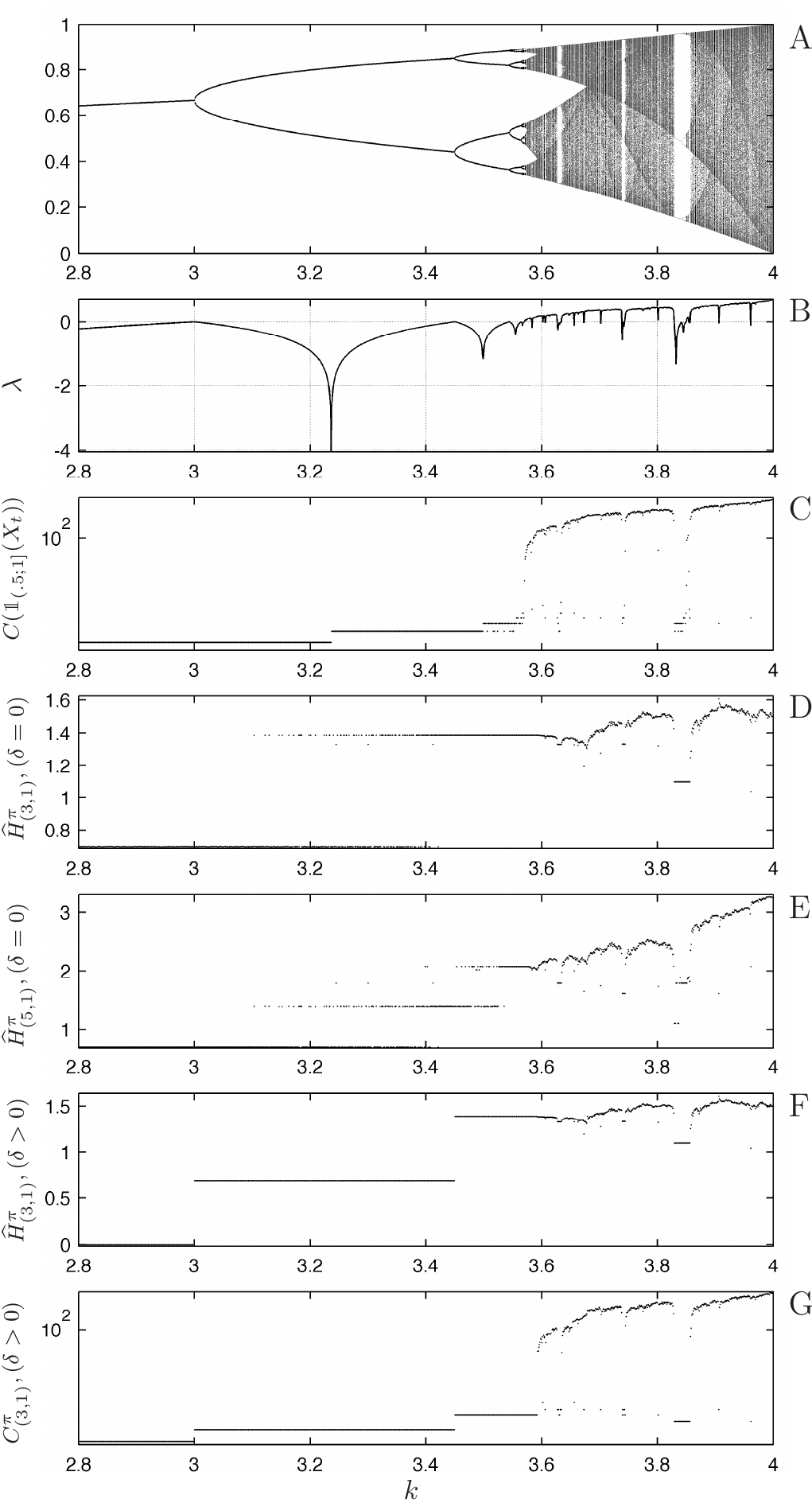}}
\caption{Characterization of  the logistic map versus $k$.   (A) The bifurcation
  diagram; \ie the  values taken by the series in the  permanent regime for each
  value  of $k$.   (B) The  Lyapunov  exponent $\lambda$.   (C) The  Lempel--Ziv
  complexity  of   the  quantized  signal   $\un_{(.5  \:  ;  \:   1]}(X_t)$  as
  in~\cite{KasSch87}. (D-F) The permutation entropy $\widehat{H}_{(d,\tau)}^\pi$
  with a  delay $\tau = 1$ when  $(d,\delta) = (3,0)$ (D),  $(d,\delta) = (5,0)$
  (E),  and $(d,\delta)  = (3,10^{-3})$  (F).  (G)  The  Lempel--Ziv permutation
  complexity $C_{(d,\tau)}^\pi$ for $(d,\tau,\delta) = (3,1,10^{-3})$.}
\label{Logistic:fig}
\end{figure}

The logistic  map has been  studied for  a long time,  and its behavior  is well
known and can be found in any textbook on chaos; \eg~\cite{AllSau96,Str94}.  Let
us just recall that when $k$  increases, it shows more and more complex regimes:
there  is an  increasing  sequence  of values  $k_{-1}  = 0  <  k_0  < \cdots  <
k_{\infty} \approx  3.56995$ such that,  if $k \in  (k_{n-1} \: ; \:  k_n]$, the
output asymptotically oscillates between $2^n$ values, a phenomenon that is well
known as {\em  bifurcations}.  For $k \ge k_\infty$, the system  is in a chaotic
(unpredictable)  regime.   Roughly  speaking,  it appears  to  behave  randomly,
although it is produced by  an elementary deterministic system. However, in this
zone, there remain some intervals, known as {\em islands of stability}, in which
the behavior  is nonchaotic.  This  briefly described behavior is  summarized in
the bifurcation diagram plotted in Figure~\ref{Logistic:fig}A.

Let  us now  study  the  regimes of  the  logistic map  versus  $k$ through  the
Lempel--Ziv permutation  complexity proposed here.   To this end, a  sequence of
size $T  = 1000$ is  drawn and only  the second half  of the sequence,  which is
assumed  to  be  in  the   permanent  regime,  is  analyzed.   The  behavior  of
$C_{(d,\tau)}^\pi$  versus $k$  is depicted  in  Figure~\ref{Logistic:fig}G, and
this is compared to the permutation entropy (Fig.~\ref{Logistic:fig}D-F), to the
Lempel--Ziv  complexity performed  on  a static  $2-$level  quantization of  the
signal  $\un_{(.5 \:  ;  \: 1]}(X_t)$,  where  $\un$ is  the indicator  function
(Fig.~\ref{Logistic:fig}C),      and     to      the      Lyapunov     exponents
(Fig.~\ref{Logistic:fig}B).       Roughly       speaking,      the      Lyapunov
exponent\footnote{For  a discrete  map  of  the type  $X_{t+1}  = f(X_t)$,  this
  coefficient  is  given  by   $\displaystyle  \lambda  =  \lim_{T  \to  \infty}
  \frac{1}{T}  \sum_{t=1}^T  \log  f'(X_t)$~\cite{AllSau96,Str94}.   Practically
  speaking,  this is  calculated for  a  large $T$.}   measures the  exponential
convergence or divergence of two  trajectories for two close initial conditions:
a positive Lyapunov exponent is a signature of chaos~\cite{AllSau96,Str94}.

The behavior of each descriptor can be interpreted as follows:
\begin{itemize}
\item {\em The  Lyapunov exponent}: This exponent clearly  describes the chaotic
  character  of  the  logistic  sequence   (when  it  is  positive)  versus  its
  non-chaotic character (when it is  negative). However, as already mentioned in
  the literature, this  is not precise enough to  distinguish different types of
  behavior in nonchaotic regimes.
\item  {\em  The  Lempel--Ziv   complexity  $c\!\left(  \left\{  \un_{(.5  \:  ;
          \:1]}(X_t) \right\}\right)$}:  As claimed by Kaspar  \& Schuster, this
  measure  is more  precise  than  the Lyapunov  exponent.   In particular,  the
  complexity is  very high  in chaotic  regimes, while it  is low  in nonchaotic
  regimes.  However, it can be seen  that the bifurcations are not detected very
  well.  This  is clearly due to  the quantization threshold.  Indeed,  for $k <
  3.237$, the system  asymptotically oscillates between two values  $> 0.5$, the
  threshold  that was  chosen  by Kaspar  \&  Schuster, which  explains why  the
  complexity fails to detect the  bifurcations.  The same phenomenon appears for
  the following  bifurcations.  Note that choosing  a threshold of  2/3 for this
  system  leads  to  the detection  of  the  first  bifurcation, but  the  other
  bifurcations remain undetected.
\item {\em  The permutation entropies $\widehat{H}_{(d,1)}^\pi$:}  In both cases
  of $d  = 3$ and $d =  5$, the permutation entropy  precisely characterizes the
  different regimes of  the logistic map.  In particular, it  is high in chaotic
  regions, while  it is low  in nonchaotic  regions; \eg as  is the case  in the
  islands  of stability.   This is  particularly true  for the  `high' embedding
  dimension; \eg  $d=5$.  Note that for  $\delta = 0$, the  first bifurcation is
  not detected here.   This is due to the small  oscillations that remain around
  the limit  value when $k \in  (2 \: ; \:  k_0]$.  The consequence  is that the
  permutation  entropy fails  to detect  the  first bifurcation,  as the  damped
  oscillatory behavior of the system for $k \in  (2 \: ; \: k_0]$ is seen in the
  same manner as the sustained oscillations of  the system when $k \in (k_0 \: ;
  \:  k_1]$.  Obviously,  the permutation  entropy ($\delta  = 0$)  detects this
  oscillatory behavior which  is inherent to the system. However,  if we are not
  really  interested in the  signal itself,  but in  its asymptotic  regime, the
  small  fluctuations can  be viewed  as perturbations.   Choosing $\delta  > 0$
  allows  the `filtering' of  these perturbations.   In this  case, even  if the
  permutation  entropy does not  characterize the  logistic sequence  itself, it
  very precisely characterizes the asymptotic regimes of the sequence, as can be
  seen in Figure~\ref{Logistic:fig}. Indeed,  in this case, the bifurcations are
  very well detected, even for `low' embedding dimensions.
\item {\em  The Lempel--Ziv permutation complexity $C_{(d,1)}^\pi$:}  At a first
  glance, this measure behaves like the permutation entropy.  In particular, the
  same effects of detection or not of the bifurcation occur if $\delta = 0$ (not
  plotted in  Figure~\ref{Logistic:fig}) or $\delta  > 0$.  Note,  however, that
  even  in  the  low  embedding  dimension, the  complexity  appears  to  better
  characterize  the  constant, oscillatory  or  chaotic  regimes.  Indeed  while
  $\widehat{H}_{(3,1)}^\pi$ is roughly constant  when the chaos appears (for $k$
  slightly $>k_\infty$), the complexity greatly increases.
\end{itemize}


\subsection{Detecting of a sudden change in a three-dimensional signal}

To illustrate how the proposed measure can outperform the permutation entropy in
assessing  the  degree  of  complexity  of  some  signals,  let  us  consider  a
multidimensional series $\vec{X}_t$ composed first of $N_c$ points issued from a
$d-$dimensional logistic series, followed by  $N_n$ points of both spatially and
temporally iid noise.  The $d-$dimensional  logistic map we have chosen here for
our purpose is described by the following equation:
\begin{equation}
\vec{X}_{t+1} = k \left( \mat{K} \, \vec{X}_t + \vec{1} \right) \odot
\vec{X}_t \odot \left( \vec{1} - \vec{X}_t \right)
\label{LogisticDD:eq}
\end{equation}
where $\vec{X}_t$ is a $d$-dimensional  vector, $\vec{1} = [1 \quad \cdots \quad
1]^t$,  $\vec{K}$  is  a $d  \times  d$  coupling  matrix,  and $\odot$  is  the
component-wise product  ($t \ge 0$). When  $\mat{K}$ is zero,  the $d$ logistics
are decoupled.  For  the opposite, when $\mat{K} = 3  \mat{P}$ with $\mat{P}$ as
the  cyclic permutation matrix  of one  place to  the left,  or when  $\mat{K} =
\vec{1} \vec{1}^t$, the map corresponds to the models proposed by Lopez-Ruiz and
Fournier-Prunaret in  the $2-$dimensional and $3-$dimensional  contexts to model
symbiotic  interactions between  species,  where parameter  $k$ represented  the
growth rate of the species~\cite{FouLop04,LopFou04}.  In both the cases of $d=2$
and $d=3$,  according to  the value of  $k$, these  maps show regular  orbits or
chaotic orbits.  We do not describe here the richness of these maps, but instead
direct the reader to~\cite{FouLop04,LopFou04}.

\begin{figure}[htbp]
\includegraphics[width=.485\textwidth]{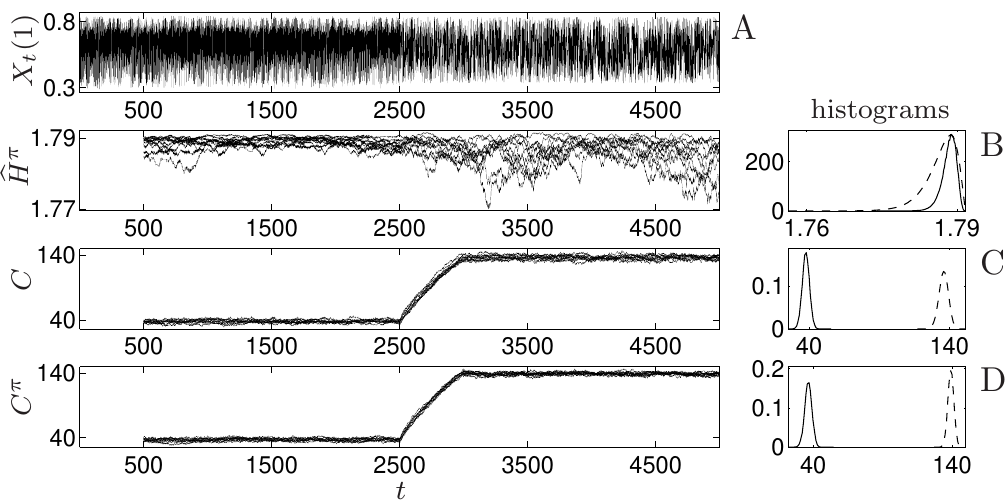}
\caption{Detection of a sudden change  in a $3-$dimensional sequence composed of
  $N_c  =  2500$ points  of  a coupled  $3-$dimensional  logistic  map given  by
  Equation~\eqref{LogisticDD:eq} followed by $N_n  = 2500$ points of pure random
  noise (uniform).   (A) A snapshot of  the first component of  such a sequence,
  with 2000  sequences then  analyzed through  a sliding window  of $N_w  = 500$
  points,  moving sample  by sample.   (B-D)  Ten snapshots  of the  permutation
  entropies  (B), the  Lempel--Ziv complexities  of a  quantized version  of the
  vectors (C), and Lempel--Ziv permutation complexity (D) are shown.  Right: the
  corresponding  histograms of  the values  taken  by the  measure, showing  the
  windows in the chaotic part (solid  line) and in the noise part (dashed line).
  The chaotic map is here strongly coupled,  with $\mat{K} = 3 \mat{P}$ and $k =
  1.01$.}
\label{Logistic3D_Strong:fig}
\end{figure}

\begin{figure}[htbp]
\includegraphics[width=.485\textwidth]{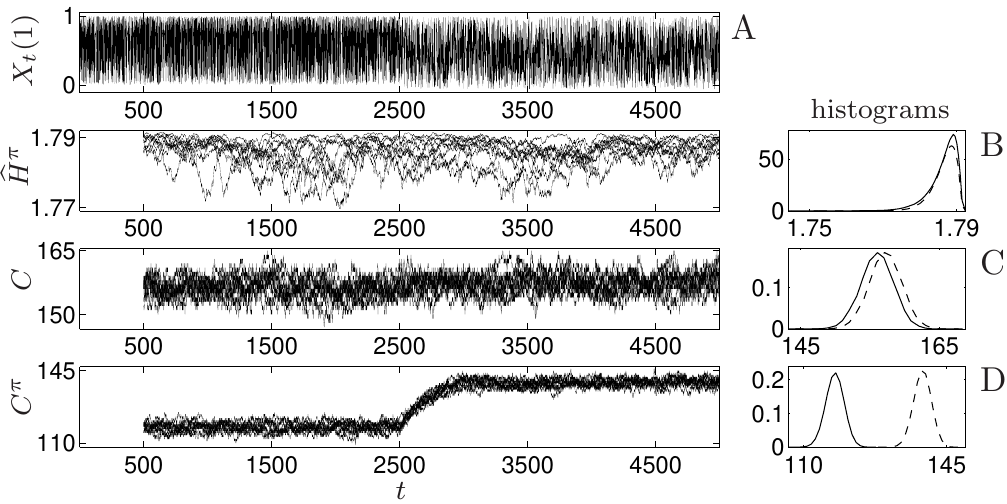}
\caption{Same  as for  Figure~\ref{Logistic3D_Strong:fig} for  a  weakly coupled
  chaotic map, with $\mat{K} = .01 \mat{P}$ and $k = 3.96$.}
\label{Logistic3D_Weak:fig}
\end{figure}

For our  purposes, we  have chosen to  study what  happens when the  $N_c$ first
points of the sequence are  generated by the $3-$dimensional map ($d=3$) showing
chaotic  behavior.  We  considered  two cases:  in  the first,  the coupling  is
$\mat{K} = 3 \mat{P}$ and $k = 1.01$; and in the second, $\mat{K} = .01 \mat{P}$
and $k  = 3.96$. In the first  case, the components are  strongly coupled, while
they are  weakly coupled in the second  case.  A snapshot of  these logistic map
sequences      followed      by       pure      noise      is      shown      in
Figures~\ref{Logistic3D_Strong:fig}A  and~\ref{Logistic3D_Weak:fig}A.  Visually,
it is relatively difficult to detect  the instant where the nature of the signal
changes.  Let us then analyze the  signal through sliding windows of size $N_w$,
moving point  by point.   In each window  of the analysis  $(\vec{X}_{t-N_w+1} ,
\ldots , \vec{X}_t)$, $t =  N_w-1, \ldots$, we evaluate the permutation entropy,
the  Lempel--Ziv  complexity  of  a  quantized version  of  the  components  (by
$\un_{[.5 \: ; \: +\infty)}$),  and the Lempel--Ziv permutation complexity.  The
results  versus   $t$  are  plotted   in  Figures~\ref{Logistic3D_Strong:fig}B-D
and~\ref{Logistic3D_Weak:fig}B-D, where 10 realizations are shown.  On the right
of  Figures~\ref{Logistic3D_Strong:fig}B-D and~\ref{Logistic3D_Weak:fig}B-D, the
corresponding  histograms  are shown\footnote{In  the  case  of the  Lempel--Ziv
  complexities, as these  values can only take on discrete  values between 2 and
  500,   their   probability  distributions   are   discrete.    By  misuse   of
  representation, we have plotted them as continuous distributions to make their
  reading  easier.}   for  the  values  taken by  each  measure  using  $4.10^6$
snapshots of the chaotic map (solid lines) and the noise (dashed lines).

In these examples, the interpretations are the following:
\begin{itemize}
\item {\em The permutation entropy}: This  index cannot detect the change in the
  nature of the signal, as can be  seen in the snapshots for both the strong and
  weak                coupling               (Figs.~\ref{Logistic3D_Strong:fig}B
  and~\ref{Logistic3D_Weak:fig}B).   This  is because,  in  these examples,  the
  patterns  obtained in  the  permutation vectors  performed  on the  components
  appear  with  similar frequencies  to  the chaotic  regime  and  in the  noise
  regime. By statistically analyzing these patterns, the dynamics underlying the
  data are lost. The difficulty in the discrimination between chaos and noise is
  also illustrated by the probabilities  taken by the values of $H^\pi$: roughly
  speaking, the probability  of error in a discrimination task  is a function of
  the surface shared by the two distributions.
\item {\em The  Lempel--Ziv complexity}: When looking at the  case of the strong
  coupling between  the components of  the logistic, the  Lempel--Ziv complexity
  performed on the  basic quantized version of the  vector clearly discriminates
  between chaos  and noise.   However, when the  components are  weakly coupled,
  this is no more the case.  This is clearly seen in the histograms that overlap
  in  the weak  coupling case  (Fig.~\ref{Logistic3D_Weak:fig}C) while  they are
  separated        in         the        strong        coupling        situation
  (Fig.~\ref{Logistic3D_Strong:fig}C).   Our interpretation  of  this effect  is
  that, in a sense, the Lempel--Ziv analyzes the components almost individually:
  in  the weak  coupling case,  it  does not  `see' that  the components  follow
  exactly  the  same dynamics  and  are,  in a  sense,  linked  by these  common
  dynamics.
\item {\em The  Lempel--Ziv permutation complexity}: In both  types of coupling,
  this  measure unambiguously detects  the change  in the  nature.  This  can be
  viewed  both in  the snapshots  and in  the probability  distributions  of the
  values    taken    by    this   measure    (Figs.~\ref{Logistic3D_Strong:fig}D
  and~\ref{Logistic3D_Weak:fig}D). Clearly, there is  no overlap between the two
  histograms,  which confirms  that  there is  no  probability of  error in  the
  discrimination between  the chaos  and noise in  this illustration.   From the
  curves,  it would  appear that  for  both cases,  the Lempel--Ziv  permutation
  entropy shows a weaker dispersion around its mean value than does the standard
  Lempel--Ziv complexity.
\end{itemize}
These illustrations show  that in spite of the power  of the permutation entropy
to discriminate between chaos and randomness, for instance, there are situations
in which  this tool fails in this  task.  Using the Lempel--Ziv  complexity of a
basic quantized version of the sequence  can be an alternative, but this remains
dependent on  the quantification.  Moreover, in  this example, when  there is no
coupling  or there  is weak  coupling  between the  components, the  permutation
vector takes into account that  the components follow exactly the same dynamics,
which is what the standard Lempel--Ziv  complexity appears not to do.  For these
interpretations,  basically,   we  believe   that  dealing  with   an  intrinsic
multidimensional  sequence,  the Lempel--Ziv  permutation  complexity should  be
preferred to the permutation entropy and the standard Lempel--Ziv complexity.


\subsection{Epileptic electroencephalogram analysis}

\begin{figure}[h]
\centerline{\includegraphics[width=.485\textwidth]{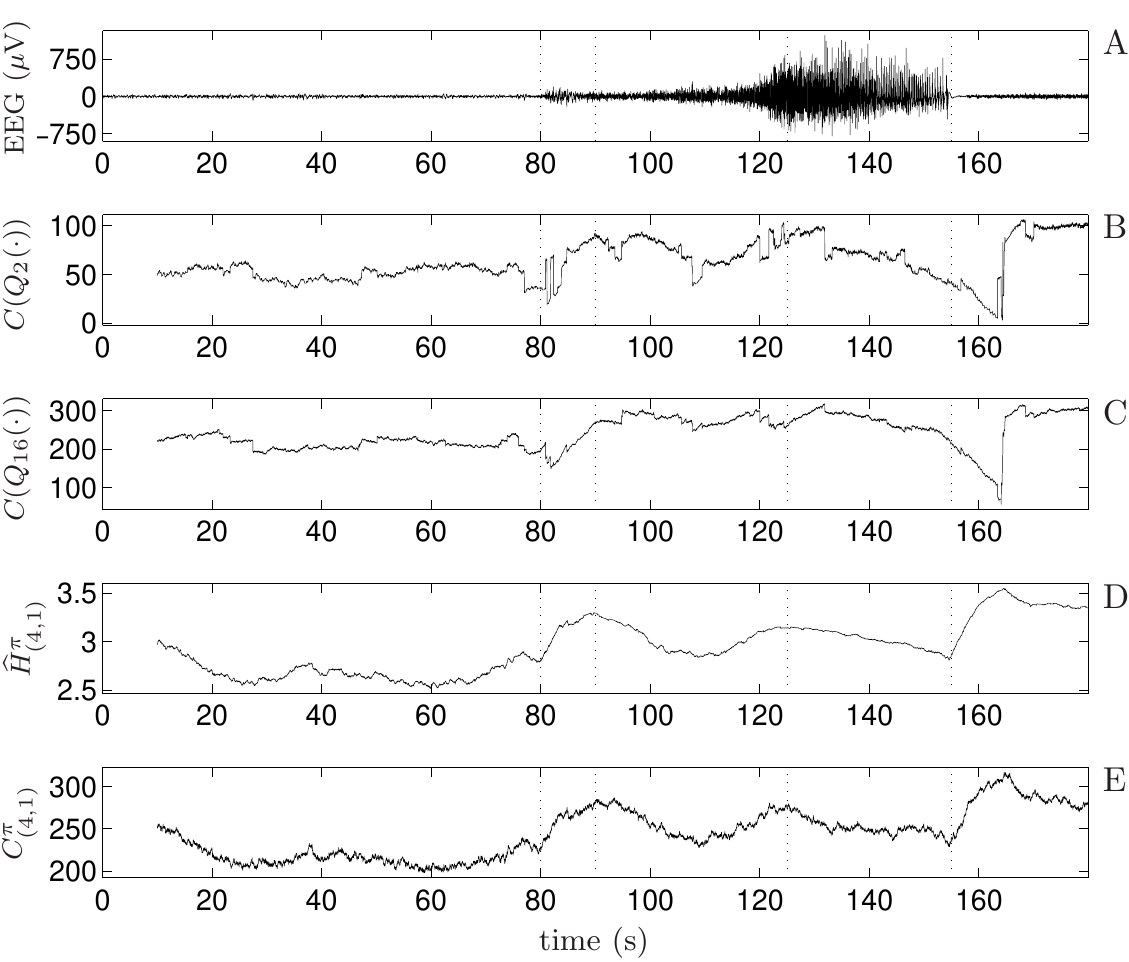}}
\caption{Electroencephalogram records of the analysis of a secondary generalized
  tonic-clonic  epileptic seizure.  The  analysis was  performed with  a sliding
  window  of 10~s  (1024  points) moving  sample  by sample.   (A) The  original
  EEG. (B, C)  The Lempel--Ziv analysis was performed  on a 2-level quantization
  (B) and a 16-level quantization (C),  and the quantizers were uniform over the
  dynamics of the  analyzed window. (D, E) For both  the permutation entropy (D)
  and the  Lempel--Ziv permutation complexity (E), the  permutation vectors were
  evaluated  from  a  reconstructed  phase-space  trajectory  with  an  embedded
  dimension $d = 4$ and a delay $\tau = 1$.  The confidence radius was chosen as
  zero.  The vertical dotted lines denote the characteristic times of $T_1, T_2,
  T_3$ and $T_4$.}
\label{EEG:fig}
\end{figure}

The electroencephalogram (EEG) signal  analyzed in this illustration corresponds
to a scalp EEG record of a secondary generalized tonic-clonic epileptic seizure,
recorded from a central right location  ($C4$) of the scalp.  This EEG record is
one of the EEGs studied by Rosso \etal in~\cite{RosBla03,RosMar06,PerLam07}.  It
was  obtained   from  a   39-year-old  female  patient   with  a   diagnosis  of
pharmaco-resistant epilepsy (temporal lobe  epilepsy), and no other accompanying
disorders.   The EEG signal  is shown  in Figure~\ref{EEG:fig}A.   The epileptic
seizure started at $T_1  = 80$~s, with a {\em discharge} of  slow waves that are
superposed by  fast waves with a  lower amplitude.  This  discharge lasts beyond
$\Delta  T  =  8$~s,  and  has  a mean  amplitude  of  100~$\mu$V.   During  the
tonic-clonic epileptic seizure, there  are very high amplitudes that contaminate
the seizure  recording, and the patient had  to be treated with  an inhibitor of
muscle responses.  After a short period, a desynchronization phase, known as the
{\em epileptic recruiting rhythm}, appears in a frequency band centered at about
10~Hz,  and it  rapidly increases  in  amplitude.  After  approximately 10~s,  a
progressive    increase   of    the   lower    frequencies    (0.5-3.5~Hz)   was
observed~\cite{GasBrou72}.  For the EEG studied here, this phase appears at $T_2
= 90$~s.  It is also possible to establish the beginning of the clonic phase, at
around $T_3 = 125$~s,  and the end of the seizure at  $T_4 = 155$~s, where there
is an abrupt decay of the signal amplitude.

The recorded  signal has  a duration  of 180~s, and  the sampling  frequency was
102.4~Hz (1024  samples/10~s) so that we  dispose of 18432  samples.  To analyze
the signal,  we again consider the  methodology proposed in  this paper; namely,
the  evaluation  of the  Lempel--Ziv  permutation  complexity.   This result  is
compared  to  that given  by  the standard  Lempel--Ziv  performed  on a  static
quantized version of the signal,  and with the permutation entropy. The analysis
was performed  with sliding windows  of size $N_w  = 1024$ points  (10~s), which
moved sample by  sample.  Here, two quantized version  are considered: a 2-level
$Q_2$ and a 16-level  $Q_{16}$, both of which are uniform over  the range of the
signal in the window of analysis.  For the permutation measures, the permutation
vectors were constructed with an embedding dimension and a delay, of $d = 4$ and
$\tau =  1$, respectively.  We chose here  a radius of confidence  of zero.  The
results are shown in Figure~\ref{EEG:fig}B-E.

The interpretations of these analyses are the following:
\begin{itemize}
\item  {\em The  Lempel--Ziv  complexity}:  For both  the  2-level and  16-level
  quantization,  this   measure  cannot  detect  any  change   in  the  analyzed
  series. Although not plotted here,  we also tested 4-level and 8-level uniform
  quantizers, which leads to the same conclusion.
\item {\em  The permutation  entropy}: In this  signal, the  permutation entropy
  detects the  appearance of  the epileptic  seizure at $T_1  = 80$~s,  which is
  visible in  the signal. The increase in  the entropy measures a  change in the
  nature of  the signal;  it is not  just a  change in amplitude,  otherwise the
  nature of the sequence of the permutation vectors would not have been changed,
  and nor would  its entropy.  Similarly, the characteristic  times $T_2 = 90$~s
  (not very visible in the signal), $T_3  = 125$~s (the clonic phase) and $T_4 =
  155$~s (end of  seizure) that are visible in the signal  are also detected (as
  decreases and an increase  in the permutation entropy, respectively). However,
  the characteristic time $T_3$ is not well detected by the permutation entropy.
\item  {\em The Lempel--Ziv  permutation complexity}:  It can  be seen  that the
  characteristic  times detected  by the  permutation entropy  are  also clearly
  detected  by  the  Lempel--Ziv  permutation  complexity.  The  shape  of  this
  complexity is very similar to  that of the permutation entropy. In particular,
  the Lempel--Ziv  permutation complexity detects  a modification of  the signal
  after the  time $T_2  = 90$ s,  a change  that is not  particularly detectable
  visually: at the  peak, the analyzed window is  completely inside the `complex
  part' of the  crisis, but the decrease indicates that  the signal becomes more
  and more  organized.  Finally,  the Lempel--Ziv permutation  complexity better
  detects the modification of the signal after  the time $T_3 = 125 $~s than the
  permutation entropy.
\end{itemize}
Note  that  both  the   permutation  entropy  and  the  Lempel--Ziv  permutation
complexity appear to indicate the appearance  of an event at time 110~s, as seen
by their  increases.  We have no  interpretation yet as to  this possible event.
Finally,  the  abrupt change  that  was  detected  by the  standard  Lempel--Ziv
complexity  at time 165~s  is only  a consequence  of the  abrupt change  in the
dynamics.

We can  see in this  example that the  measure of complexity introduced  in this
paper increases steeply  and very precisely in time when  the patient starts the
seizure, and even  more, it can detect the different  states of the tonic-clonic
epileptic seizure.  Note also the high level of the complexity at the end of the
signal compared to that at the  beginning.  This level indicates that the signal
remains  `disorganized'. A possible  interpretation of  such high  complexity is
that  even if  the  epileptic  sequence is  apparently  ended, complex  activity
remains  consequent to  the crisis.   A longer  post-epilepsy sequence  would be
needed  to verify whether  the complexity  decreases to  the low  value observed
before the crisis.

As this  signal serves essentially as an  illustration, and as our  goal here is
not to carry out  deep EEG analyses, we will not go  further with this analysis.
We    also   do    not   compare    our    result   here    to   those    obtain
in~\cite{RosBla03,RosMar06,PerLam07}, which merits a study in itself.


\section{Discussion}
\label{Conclusion:sec}

Data  analysis has  a long  history and  still gives  rise to  a huge  amount of
research. Among the  challenges, especially for the analysis  of natural signals
such as  biomedical signals,  there is  the need to  characterize the  degree of
organization or  the degree  of complexity of  signal sequences, the  problem of
detecting  sudden sequence  changes that  are not  detectable visually,  and the
problem of characterization of the nature of specific changes in a sequence. The
literature  on information  theory on  the one  hand, and  on  dynamical systems
analysis on  the other,  provides an  important number of  tools and  methods to
solve these challenges.

In this paper, we propose a tool  that mixes two very well known approaches: the
permutation entropy and  the Lempel--Ziv complexity. The idea is  to try to take
the advantage  of both of these  approaches, the first of  which is statistical,
and the second of which is deterministic.

The Lempel--Ziv complexity  has long been known and  was initially introduced in
the compression  domain.  However,  it has  been shown to  be powerful  for data
analysis.   On the  other hand,  the permutation  entropy allows  a part  of the
dynamics of  a signal  underlying data to  be captured  when it is  performed on
reconstructed phase--space signals.   Moreover, in some sense, it  is based on a
kind of quantization of the data, by considering only the tendencies rather than
the values of the sequence. From this last, it appears natural to quantize data,
as has  been done via the permutation  vectors of a vector  sequence (natural or
reconstructed)  followed by  the  evaluation of  the  the complexity  of such  a
quantized sequence.   The association  of these two  approaches has  here `given
birth' to what we have named the Lempel--Ziv permutation complexity, which is at
the heart of our proposal.

In  this paper, in  particular, we  have shown  how the  Lempel--Ziv permutation
complexity of  a sequence  can precisely capture  the degree of  organization of
such series.   When dealing with  scalar sequences, the  Lempel--Ziv permutation
complexity appears to give similar  results to those of the permutation entropy,
even  if one measure  is statistical  while the  other is  purely deterministic.
However,   when  dealing  with   intrinsic  multidimensional   signals,  without
procedures  of  phase--space  reconstruction,   the  entropy  performed  on  the
permutation vectors built from the  vector sequences cannot capture the dynamics
that underlie the data.  Indeed,  the calculation of the frequency of occurrence
of such  permutation vectors  is then a  point-by-point analysis, and  the links
between successive  points are lost.  Conversely, as  the Lempel--Ziv complexity
aims  to  detect  regularities  in  a  sequence by  analyzing  how  the  symbols
(numerical  scalar samples, vectors,  or any  kind of  symbol) can  be predicted
algorithmically from the  past symbols, it captures the  dynamics of the signal.
Doing  this analysis  for the  permutation vector  sequences allows  the natural
solving  of the  question of  quantization of  the data,  as by  definition, the
Lempel--Ziv  complexity works  with sequences  of  symbols lying  on a  discrete
finite  size  alphabet.  As  shown  in our  illustration,  we  can imagine  many
situations for which the Lempel--Ziv permutation complexity can capture a degree
of organization,  while the permutation  entropy fails, especially  when dealing
with multidimensional signals; \ie without phase-space (re)construction.


\begin{acknowledgement}
  S. Zozor is grateful to the R\'egion Rh\^one-Alpes (France) for the grant that
  enabled this work. D. Mateos is a Fellowship holder of SeCyT, UNC.
\end{acknowledgement}


\appendix

\section{Technical details}
\label{TechnicalDetails:app}

Before detailing a  possible practical implementation, we should  point out that
when two components of a vector  are equal, an ambiguity remains when performing
the permutation procedure.  Such a  situation appears with a probability of zero
for continuous  state iid  random sequences,  but it can  appear in  constant or
periodic sequences,  for instance.  To avoid  such an ambiguity,  Bandt \& Pompe
proposed  to add  a small  perturbation to  the values,  which is  equivalent to
choosing randomly the `smallest' value  between two equal values.  For instance,
in the  example of  a constant  sequence, in this  way, the  permutation vectors
reflect only  the behavior  of the perturbation,  and thus both  the permutation
entropy and the  Lempel--Ziv permutation complexity are of the  noise and not of
the signal  under analysis.   To overcome  such a difficulty,  we chose  here to
consider that  the `smallest' of  two equal values  as the `oldest' one,  as has
also been  done in  the literature.  In  the example  of a constant  signal, the
sequence of permutation vectors will be constant, which can then capture the low
complexity of the sequence.

Conversely,  an  observed  sequence can  be  corrupted  by  a low  noise.   This
corrupting noise  can hide the complexity  of the sequence  when the permutation
vectors are evaluated.  The example  of a constant signal again illustrates such
an impact of the noise. To  counteract perturbations, a way to denoise or filter
the observed sequence can consist of choosing a value $\delta \ge 0$ so that for
two components  $Y(i)$ and $Y(j)$ of  a (phase-space) vector, if  $\left| Y(i) -
  Y(j) \right| \le \delta$ then $Y(i)$  and $Y(j)$ are interpreted as equal.  In
a  sense, $\delta$ is  a {\em  radius of  confidence} in  the measured  data. If
$\delta = 0$,  this means that we have perfect confidence  in the measured data,
while  for $\delta  > 0$  we  take into  account possible  perturbations in  the
measures.  In other words, $\delta$ can  be chosen to be equal to the resolution
of the acquisition.

Practically, to evaluated $C_{d,\tau}^\pi(X_t)$, and to avoid two passes through
the sequence,  this can be done  recursively, by alternating  the calculation of
the permutation vectors and the up-dating of the complexity:
\begin{enumerate}[Step 1.]\setcounter{enumi}{-1}
\item   Construction   of   the   first  $d-$dimensional   vector   $\vec{Y}   =
  \vec{Y}_t^{(d,\tau)}$   and  evaluation  of   the  first   permutation  vector
  $\vec{\Pi}_t  = \vec{\Pi}_{\vec{Y}}$,  $t =  0$; storage  of  this permutation
  vector  in   a  stack,  and   initialization  of  the   Lempel--Ziv  algorithm
  (implicitly, the first production step).
\item\label{BP:it}  $t \leftarrow  t+1$:  replacement of  $\vec{Y}$  by the  new
  vector  of   the  trajectory,  evaluation   of  the  new   permutation  vector
  $\vec{\Pi}_t$ to be stored in the stack.
\item\label{LZ:it}   Up-dating  of   the  Lempel--Ziv   complexity   using  this
  permutation vector, and go to step~\ref{BP:it}.
\end{enumerate}
In  the  case  where $\tau  =  1$,  the  evaluation  of the  permutation  vector
$\vec{\Pi}_t$ at time $t$ can be simplified by using $\vec{\Pi}_{t-1}$.  Indeed,
in   the   constructed   trajectory    vector   $\vec{Y}$,   the   first   point
$X_{\mathrm{out}} =  Y(0)$ disappears, the  other $d-1$ components  are shifted,
and the next point of the scalar sequence $X_t$ appears as the last component of
$\vec{Y}$.  The  permutation of component $i$  (previously $i+1, i  = 1, \ldots,
d-1$) changes only if either $X_t \ge Y(i)$ and $X_{\mathrm{out}} \le Y(i)$ (the
rank  decreases)  or  $X_t <  Y(i)$  and  $X_{\mathrm{out}}  > Y(i)$  (the  rank
increases).  This  up-dating of the  rank can thus  be made with $d$  doublet of
comparisons (seeking also the rank of the new point $X_t$).
 
For  the Lempel--Ziv  complexity,  when  beginning a  new  production step,  the
algorithm  of~\cite{KasSch87} consists  of testing  all  of the  letters of  the
already  constructed history  as possible  pointers  of a  production step,  and
retaining the letter that gives the greatest production step: this pointer gives
what is then called an {\em exhaustive} production step.

\begin{figure*}
\centerline{\includegraphics[height=.8\textheight]{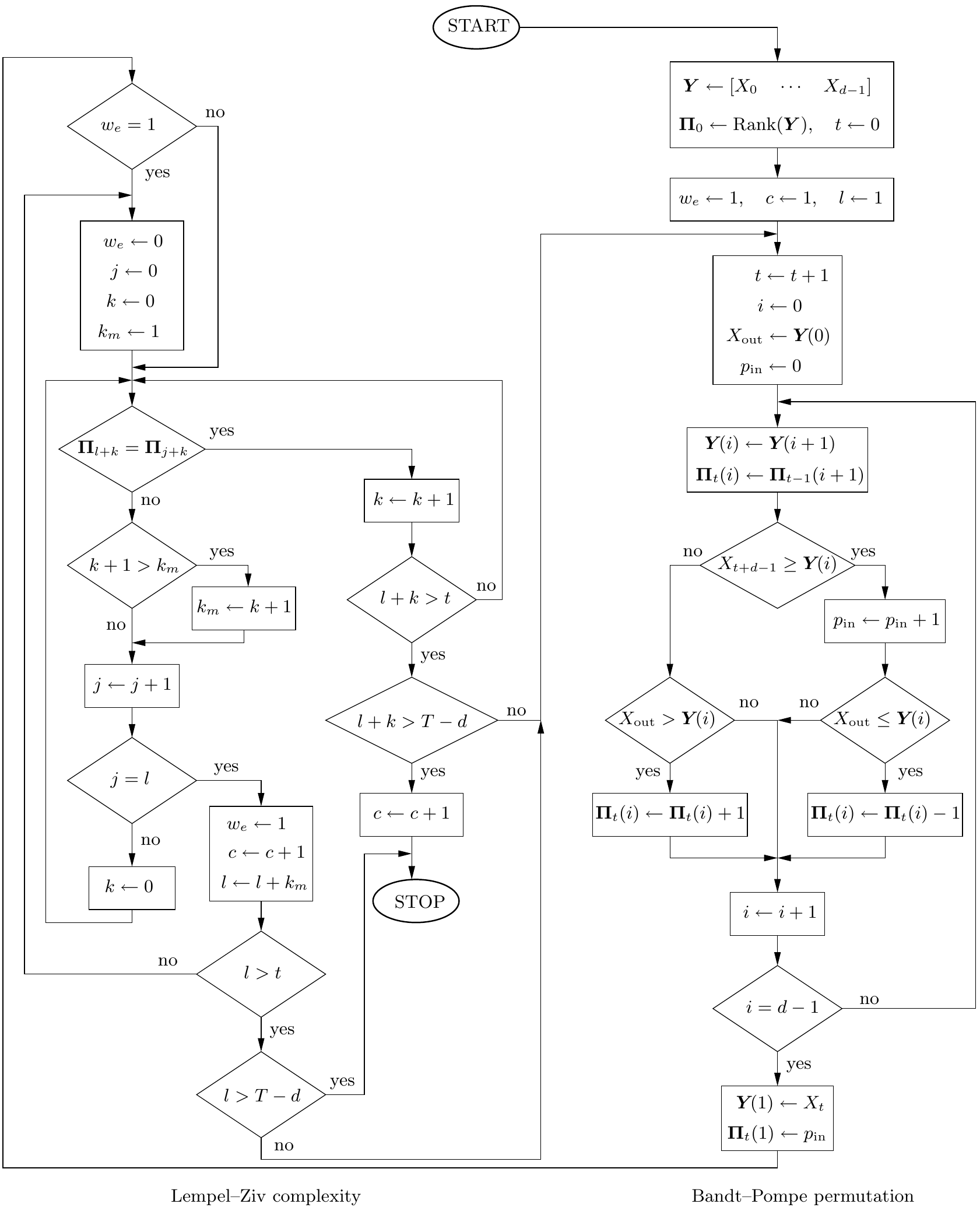}}
\caption{Diagram flow  of the  algorithm evaluating the  Lempel--Ziv permutation
  complexity $C^\pi_{d,\tau}$ for  a scalar sequence.  In this  diagram, $\tau =
  1$ (see text for the extension to any $\tau$), and the size of the sequence is
  denoted as  $T$. $w_e$  marks when  a word is  exhaustive or  not, $l$  is the
  beginning of an  exhaustive word, $j$ is the tested pointer,  and $k_m$ is the
  size of the current exhaustive word \cite{LemZiv76,KasSch87}.}
\label{BP_LZ_Online:fig}
\end{figure*}

The global recursive algorithm is described  in detail by the diagram flow shown
in Figure~\ref{BP_LZ_Online:fig}; in this simple  case, $\tau = 1$.  For $\tau >
1$,  the same  scheme  holds, except  that we  have  to first  store the  $\tau$
permutation  vectors,  then store  the  $\tau$  vectors  $\vec{Y}$, let  us  say
$\vec{Y}_0, \ldots , \vec{Y}_{\tau-1}$,  and use both $\vec{Y}_{t \, \mathrm{\bf
    mod}   \,   \tau}$    and   $\vec{R}_{t-\tau}$   to   recursively   evaluate
$\vec{\Pi}_t$. For a  non-zero radius of confidence, in  the algorithm described
in Figure~\ref{BP_LZ_Online:fig}, $x > y$  (and respectively, $x \ge y$) is then
replaced  by $x  > y  + \delta$  (respectively, $x  \ge y+\delta$)  and $x  < y$
(respectively, $x \le y$) by $x < y - \delta$ (respectively, $x \le y-\delta$).

Note    that    there   are    various    fast    algorithms    that   rank    a
vector~\cite{Hoa62,Sed78}.  In  general, these work  by recursively partitioning
the  points  to be  ranked  in  a partially  ordered  manner  (through a  tree),
performing a brute-force sorting in the  last partitions, and coming back to the
overall  ensemble.  In  general, the  computational cost  is in  $O(d  \log d)$,
instead of $O(d^2)$ for a totally  brute force method. Such approaches should be
used in our  algorithm, using the partitions at step $t-1$  to determine that at
step  $t$,  expecting  a computational  cost  in  $O(\log  d)$ instead  of  $d$.
However,  in  practice,  the  Bandt--Pompe  entropy (and  here  the  Lempel--Ziv
permutation complexity) is studied in  low dimensions, so that the computational
cost of a  brute force approach is relatively close to  that of fast approaches.
Thus, we  will not  go deeper  into such possible  improvements of  the proposed
algorithm.

Finally,  note  that  contrary  to  the  permutation  entropy,  the  Lempel--Ziv
complexities can be evaluated online, \ie up-dated acquisition by acquisition.

\bibliographystyle{unsrt}
\bibliography{BP_CLZ_R1}

\end{document}